\documentclass[sigconf]{acmart}

\usepackage{booktabs} 
\usepackage{threeparttable}
\usepackage{multirow}
\usepackage{booktabs}
\usepackage{tabulary}

\usepackage{subcaption}
\usepackage{color}
\usepackage{url}
\usepackage{wrapfig}
\usepackage{dblfloatfix}
\usepackage{placeins}
\usepackage{float}
\usepackage{afterpage}

\begin{document}
\title{Cascading Behavior in Yelp Reviews}

\author{Muhammad Raza Khan}
\authornote{Under review}
\affiliation{%
  \institution{University of California, Berkeley}
}
\email{mraza@berkeley.edu}

\newenvironment{packed_item}{
	\begin{itemize}
		\setlength{\itemsep}{1pt}
		\setlength{\parskip}{0pt}
		\setlength{\parsep}{0pt}
	}{\end{itemize}}

\newenvironment{packed_item2}{
	\begin{itemize}
		\setlength{\itemsep}{3pt}
		\setlength{\parskip}{0pt}
		\setlength{\parsep}{0pt}
	}{\end{itemize}}

\newenvironment{packed_enum}{
	\begin{enumerate}
		\setlength{\itemsep}{1pt}
		\setlength{\parskip}{0pt}
		\setlength{\parsep}{0pt}
	}{\end{enumerate}}

\newenvironment{packed_enum2}{
	\begin{enumerate}
		\setlength{\itemsep}{3pt}
		\setlength{\parskip}{0pt}
		\setlength{\parsep}{0pt}
	}{\end{enumerate}}

\begin{abstract}

	Social media has changed the landscape of marketing and consumer research as the adoption and promotion of businesses is becoming more and more dependent on how the customers are interacting and feeling about the business on platforms like Facebook, Twitter, Yelp etc. Social review websites like Yelp have become an important source of information about different businesses. Social influence on these online platforms can result in individuals adopting or promoting ideas and actions resulting in information cascades. Research on information cascades have been gaining popularity over the last few years but most of the research has been focused on platforms like Twitter and Facebook. Research on the adoption or promotion of product using cascades can help determine important latent patterns of social influence. 

In this work, we have analyzed the spread of information i.e. cascades in Yelp across different cities in Europe and North America. We have enumerated and analyzed different cascade topologies that occur in the Yelp social networks. Some of our significant findings include the presence of a significant number of cascades in Yelp reviews indicating the importance of social influence, heavy tailed distribution of cascades and possibility to accurately predict the size of cascades on the basis of initial reviews. In addition, we have also found that the characteristics of the non root nodes and the non root reviews are much more important type of feature as compared to the properties of the root nodes and root reviews of the cascade.

These findings can help social scientists to analyze customer behavior across different cities in a much more systematic way. Furthermore, it can also help the businesses in a city to figure out different consumer trends and hence improve their processes and offerings.

\end{abstract}

%
%

\ccsdesc[300]{Human-centered computing~Social network analysis}%
%
%

\maketitle

\section{Introduction}
\label{sec:intro}

\begin{table*}[htb]\centering
\begin{threeparttable}
\renewcommand{\arraystretch}{1.1}
\footnotesize
\begin{tabular}{l*{6}{c}}
\addlinespace
\toprule
\multirow{2}{*}{\textbf{City}} & \multirow{2}{*}{\textbf{No of businesses}} & \multirow{2}{*}{\textbf{No of reviews}} & \multirow{2}{*}{\textbf{No of users}} &\multirow{2}{*}{\textbf{Total no of cascades}}&\multirow{2}{3cm}{\textbf{CL90$=$Cascades Length\textsubscript{90}}}&\multirow{2}{3cm}{\textbf{Long Cascades\\ (Length $>$ CL90)}}\\
\\
\midrule
Karlsruhe  & 2,905  & 33,888  & 12,266 & 8 & 2.2 &1\\
Edinburgh  & 3,539  & 50,781  & 12,261 & 300 &6.0&28\\
Montreal  & 6,668  & 131,927  & 52,160 & 561&5.0&46\\
Pittsburgh  & 8,091  & 212,107  & 76,577 & 823&8.0&63\\
Charlotte  & 10,177  & 282,884  & 99,345 & 896&6.0&69\\
Madison  & 3,899  & 102,140  & 38,698 &449&7.0&34\\
Urbana Champaign  & 1,556  & 33,176  & 15,431 &185&4.1&15\\
Cleveland  & 9,966  & 228,990  & 81,176 &108&2.0&6\\
Waterloo  & 24,507  & 585,119  & 126,970 &462&2.0&24\\
Las Vegas  & 28,214  & 1,846,944  & 627,609 &5,665&4.0&312\\
Pheonix  & 43,492  & 1,556,500  & 429,634 &3,267&3.0&188\\
\bottomrule
\end{tabular}
\begin{tablenotes}[normal,flushleft]
\scriptsize
\item \emph{Notes}: CL90 indicates the 90th percentile for the length of all the cascades in a particular city
\item Last column indicates the number of long cascades in each of the city
\end{tablenotes}
\caption{Summary statistics by city\label{tab:sumstats}}
\end{threeparttable}
\end{table*}

Social media websites like Facebook and Twitter etc. have become an important source of information to analyze human behavior in different circumstances. Before the popularity of these social media websites, the research on human behavior in many cases was constrained by the lack of availability of data. However, these social media websites have become the biggest observatory of human behavior, as a result of which quite a significant number of research projects on human behavior analysis have employed social media data. People express their personal feelings and concerns and at the same time review businesses and products on these social media websites. On the other end, businesses and companies have been using these social media outlets to promote their products, customer service and quality. The way customers deal with these promotions or express their views about these companies can tell a lot of information about these businesses. The research on social media analytics is primarily dominated by the work correlating business performances with the reviews, however the impact of the reviews of the individuals on others has been a relatively less popular theme. The impact of the reviews of an individual activity in the social network over other individuals' activity is the broader theme of this work.
 
Social networks impact human behavior at different levels in different contexts and it has been difficult to estimate the social influence on individual's behavior specially when it comes to buying something or dealing with a business. To quote Subramani and Rajagopalan \cite{Subramani:influence_viralmarketing:2003}, "there needs to be a greater understanding of the contexts in which viral marketing strategy works and the characteristics of products and services for which it is most effective. ... What is missing is an analysis of viral marketing that highlights systematic patterns in the nature of knowledge-sharing and persuasion by influencers and responses by recipients in online social networks". Part of the problem in effectively analyzing the influence of social networks has been the lack of availability of data, however the growing research on Facebook and Twitter data indicates that this is becoming less of a problem. The other half of the problem is that the impact of users on other users can be a complex function of different latent features. The influence of influencers on others can be different even with in a social network depending on the geographic location or other characteristics of the users.
Information cascades are a common phenomena in the social networks in which the user adopt a new idea or perform an action under the influence or recommendation of other users. Information cascades are an ideal tool to analyze the impact of social influence. Furthermore, they can also help in the analysis of important factors behind social influence in different contexts. In this work, we have analyzed  millions of reviews over the social review website Yelp from million of users across different cities in Europe and North America. 
\subsection{Present Work}
Using the data officially provided by Yelp\footnote{\url{https://www.yelp.com/dataset_challenge/}}, we explore the following questions in this research paper.

\begin{itemize}
	\item What kind of information cascades frequently arise in Yelp reviews across different cities?
	\item What are the significant features of cascades across different cities?
	\item Given information about a few early reviews, can one predict the longevity of the influence? (i.e. Whether these reviews are a part of a big or small cascade?)
\end{itemize}
Using Yelp reviews and Yelp network data from different cities we have tried to have a systematic understanding of patterns of influence across different cities exploring what sort of generic patterns can be deduced. 

\subsection{Summary of results}
There are four substantive and one methodological contributions of this study. Substantively, we (1) develop a richer understanding of what drives the diffusion of information in Yelp reviews; (2) construct a supervised learner that can predict, to varying degrees of accuracy depending on the city context, the likelihood than a cascade is going to be a big one or short one; (3) analyze the structural properties of the cascades across different cities and indirectly analyze the properties of the underlying social network; and (4) analyze the features that can help in predictability of the virality of the cascade.
To our knowledge, the analysis of information diffusion at this scale across different cultural contexts has not been performed earlier.

Methodologically, we have developed a framework to (1) enumerate and analyze cascades in a scalable and efficient manner; (2) extract the features of cascades (3) and predict whether the cascade is a long one or short one. 

The analysis of cascades can be beneficial for both the business owners and the review websites (Yelp in this case) as it can unearth latent trends of consumer behaviors in different cultures which can in turn help the business owners to improve their processes.

\section{Related Work}
\label{sec:related}
Our work builds on several distinct strands in the academic literature. The first is concerned with the general diffusion of information in social networks. This has been a popular topic of research over the last decade.\cite{cha:characterizing_flickr:2008} et al. have tried to analyze the cascading behavior in Flickr and their key finding has been that growth of cascades in Flickr can be much faster as compared to some epidemics like infectious diseases indicating that social networks can be a very effective medium for transmission of information. \cite{dow:anatomy_facebook_cascades:2013} et al. have tried to compare and contrast the growth of two cascades on Facebook. Consistent with other studies they have found out that small proportion of content generates cascades of non-trivial size and depth. However, their finding that different cascades can reach the same level of popularity through different network topologies and different source and chain characteristics is of special relevance for us. This is because, we also wanted to find out what are the frequent cascades shapes in different contexts and what part do the features of the source node and the features of the chains play in the growth of the cascade.

\cite{leskovec:cascades_blog:2007} have tried to analyze the impact of blogs on each other. The most relevant literature to our work is the work on patterns of influence in recommendation networks \cite{leskovec:patterns_recnet:2006}, \cite{leskovec:dynamics_viral:2007}. In these works, the authors have tried to analyze the patterns of social influence in recommendation networks. Their work and our work is quite similar in methods but we want to find the patterns of influence specifically in the Yelp social network. Futhermore, we want to identify important determinants for the growth of cascades in social networks. Our work also confirms the finding of the \cite{cheng:be_predicted:2014} as we also find out that cascades can be predicted quite accurately using the information about the seed nodes and initial nodes of the cascades.

Another set of literature related to this research includes the work that explores the impact of different structural characteristics of social networks to model product adoption and diffusion of information. Bakshy et al. found that the adoption rate in the Second Life social network increases as the number of adopting friends increases and the effect of the number of adopting friends depends on the connectivity of the individual user \cite{Bakshy:social_influence:2009}. In another related strand of literature, the researchers have tried to analyze the impact of social networks on product adoption across different social networks (For instance, Khan et al.  have tried to the explore the impact of different features like the alter of an ego node for the purpose of product adoption in developing countries \cite{Khan:product_adoption:2016}) and there has been mixed evidence about the impact of social network. Though the consensus is that the social networks generally do play a role in product adoption, the extent of impact of these social networks varies. Similarly, Ugander et al. found that probability of contagion is more tightly connected to the structural components in an individual's network than the actual size of the network \cite{ugander:structural:2012} while Romero et al.  have tried to analyze the interplay between social and topical structure for predicting the growth of hashtags \cite{romero:interplay:2013}. Their work is related to our research as we also try to analyze the impact of different features on the growth of the cascades.\\
There has been limited work on diffusion of information in Yelp reviews. For instance Gee et al. \cite{gee:herding_yelp:2011} have tried to explore whether a modified herding model better explains the existence of cascades in yelp or not. Though the dataset that they have been using is quite similar to what we are doing but we are more focused on the features that can help predicting the growth of the cascades across different cultures.\\
One distinguishing feature of this study is that we have analyzed the spread of cascades on Yelp which can be slightly different from the platforms like Facebook and Twitter where the users may not have to purchase any thing.  
To conclude, this work is different from other works as it deals with the Yelp social network and is focused on finding systematic patterns of influence across different cities. Relative to the studies mentioned in this section, our study confirms the existence and heavy tailed nature of cascades. But by focusing on the different type of features that can help in distinguishing between long and short cascades we have tried to narrow down on the factors that determine the popularity or virality of cascades generically across different cities.

\section{Data and Context}
\label{sec:data}
\begin{figure}[!t]%
	\centering
	\includegraphics[width=0.5\textwidth]{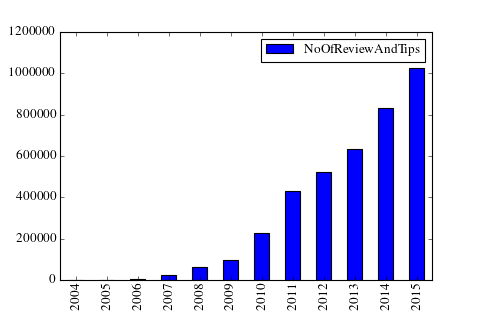}%
	\caption{No of reviews and tips per year\label{fig:review_counts}}%
\end{figure}

For this particular study, we have been using the data provided by the Yelp under its Dataset challenge initiative\footnote{\url{https://www.yelp.com/dataset_challenge/}}. As our basic aim in this research was to see the underlying patterns of social influence across different cities we decided to use the data both from the Round 8 and Round 9 of the challenge. The data in total consists of around 6 million reviews and 1.5 million tips for more than a million users and 150 thousand users across 11 different cities in the Europe and North America. In addition, to the reviews and tips the dataset also contains the business attributes like hours, parking availability and ambiance etc. as well.
Though the dataset is quite big in total but the number of reviews is quite low in the earlier years as shown in the Figure \ref{fig:review_counts}. For the purpose of this research, we have focused on cascades resulting from reviews and tips for the year 2015, though a similar analysis can be easily done for any other year provided sufficient data is present. Some of the summary statistics for each of the cities in our analysis can be found in the Table \ref{tab:sumstats}.

To find out cascades in the data we had to find out the reviews and tips that result in future review and tips. Though the behavior of the person deciding to deal with a business can be influenced by many different factors like visibility of the reviews, explicit referral from an acquaintance; we are adopting a more strict approach such that only the friends of a user can be influenced by the reviews or tips of the users. In other words, an edge $(u,v)$ will be part of a cascade for a business $b$ if the user $u$ writes a review or tip at time $t$ and the user $v$ write a review or tip at the time $t'$ such that $t' > t$ and $v \in N(u)$, where $N(u)$ specifies the neighboring nodes of $u$. All the edges that specify this property will constitute a cascade for the business $b$.

To frame the problem, as a prediction task, we calculate the length of each cascade for a business as the number of nodes or users who participate in the cascade either by writing a review or tip for the business. Furthermore, we categorized the cascades as long or short cascades as explained below
\begin{packed_item2}
	\item \textbf{Long Cascades: } If the length of a cascade is greater than or equal to the 90th percentile of the length of all the cascades of the city. The 90th percentile of the lengths of all the cascades for a city is shown in the second last column of the table \ref{tab:sumstats}.
	\item \textbf{Short Cascades: } If the length of a cascade is less than the 90th percentile of the length of all the cascades of the city.
	
\end{packed_item2}

As our primary interest in this work was to compare long vs short cascades in a city, we had to eliminate Karlsruhe, Waterloo and Cleveland from our analysis as most of the cascades in these cities were short cascades. However, given enough data our models and analysis can be easily extended to these cities as well.

\section{Structural Analysis of Cascades}
\label{sec:distribution}

\begin{figure*}[!t]
	\begin{minipage}{1.6in}
		\includegraphics[width=\linewidth]{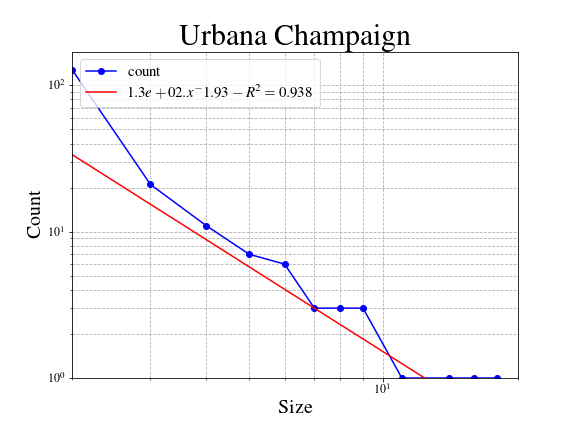}
	\end{minipage}
	\begin{minipage}{1.6in}
		\includegraphics[width=\linewidth]{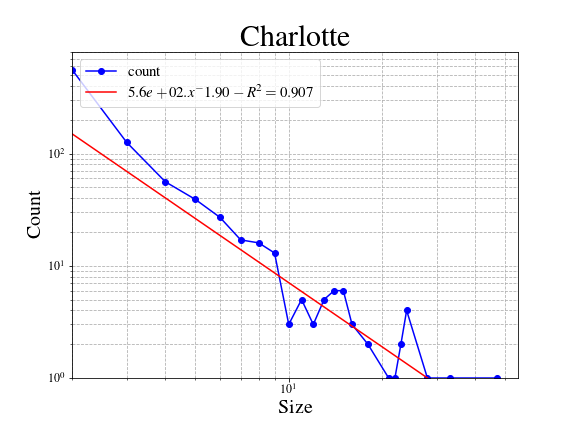}
	\end{minipage}
	\begin{minipage}{1.6in}
		\includegraphics[width=\linewidth]{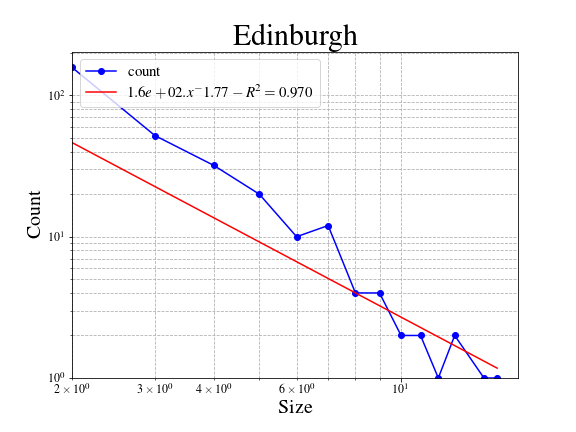}
	\end{minipage}
	\begin{minipage}{1.6in}
		\includegraphics[width=\linewidth]{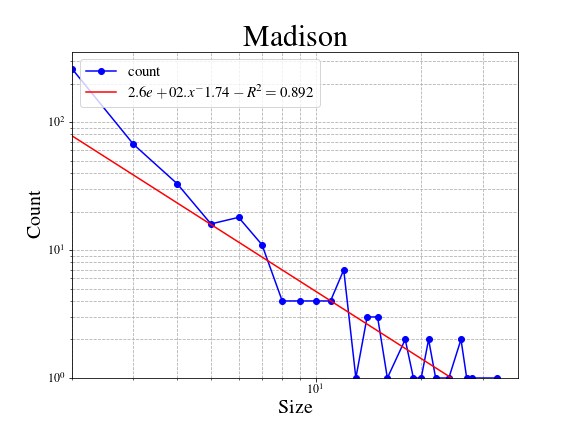}
	\end{minipage}
	\begin{minipage}{1.6in}
		\includegraphics[width=\linewidth]{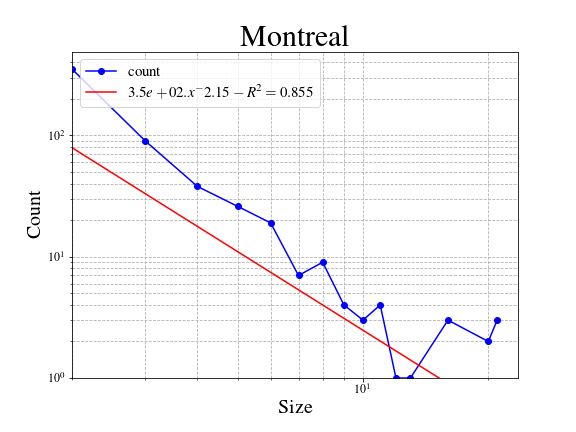}
	\end{minipage}
	\begin{minipage}{1.6in}
		\includegraphics[width=\linewidth]{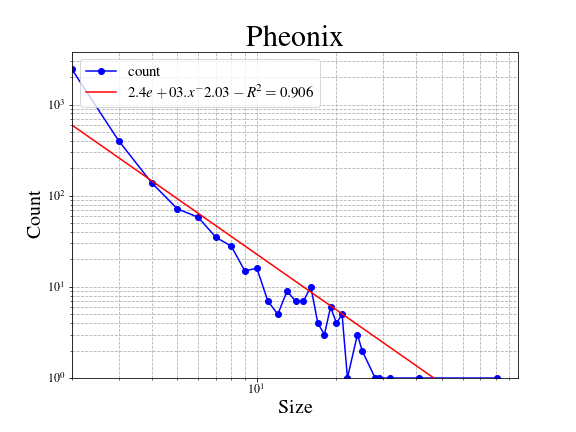}
	\end{minipage}
	\begin{minipage}{1.6in}
		\includegraphics[width=\linewidth]{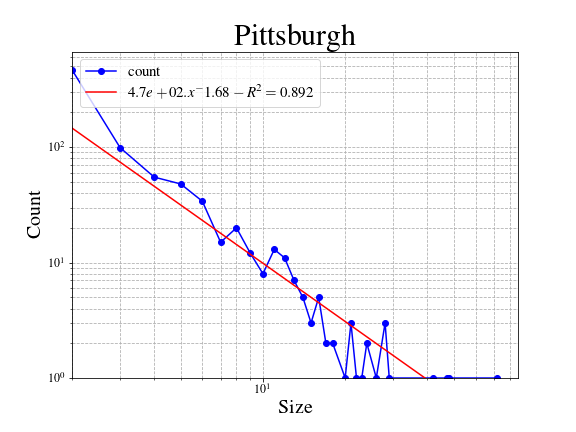}
	\end{minipage}
	\begin{minipage}{1.6in}
		\includegraphics[width=\linewidth]{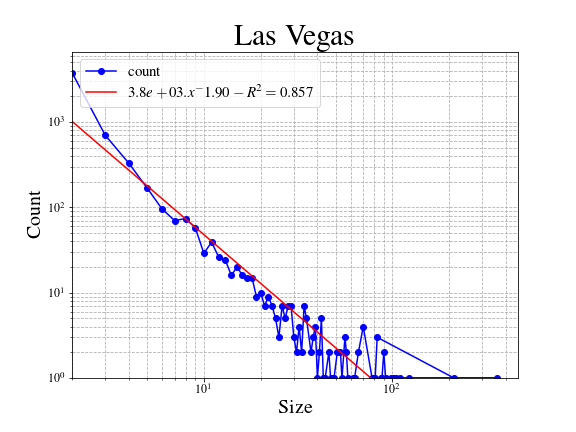}
	\end{minipage}	
	\caption{Distribution of cascades across cities\label{fig:dist}}
\end{figure*}

\subsection*{Size Distribution of Cascades}

As the first step in our analysis, we analyze the distribution of the size of cascades in different cities. The distribution of size of cascades for different cities is shown in Fig. \ref{fig:dist}. It is obvious from the Fig. \ref{fig:dist} that cascades in all of the cities follow a heavy tailed distribution. Most of the cascades are small but large cascades do happen. Furthermore, this figure also shows that the cascade distribution across different cities has a power-law exponent in the range $-2.15$ and $-1.68$. Las Vegas and Pheonix have the highest number of large cascades but large cascades do exist for other cities as well.
\begin{table*}[!htb]
	\centering
\resizebox{\textwidth}{!}{
\begin{tabular}{l|c||c|c||c|c||c|c||c|c||c|c||c|c||c|c||c|c||c|c||c|c||c|c}
	\hline
	 \multirow{2}{*}{}&\multirow{2}{*}{Graph} & \multicolumn{2}{c}{Karlsruhe} & \multicolumn{2}{c}{Edinburgh} & \multicolumn{2}{c}{Montreal} &\multicolumn{2}{c}{Pittsburgh} & \multicolumn{2}{c}{Charlotte} & \multicolumn{2}{c}{Madison} & \multicolumn{2}{c}{\parbox{1.5cm}{Urbana Champaign}}&\multicolumn{2}{c}{Cleveland}&\multicolumn{2}{c}{Waterloo}&\multicolumn{2}{c}{Las Vegas}&\multicolumn{2}{c}{Phoenix}\\
	  &&   R &   FP&   R&   FP &   R&   FP &   R & FP &   R &  FP &   R &   FP &   R&   FP &   R &   FP &   R&   FP &   R&   FP &   R &   FP\\
\hline
G1&\includegraphics[scale=0.3]{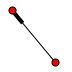}  &                1 &                  87.5 &                1 &                 51    &               1 &                62.03 &                 1 &                  55.53 &                1 &                 61.05 &              1 &               57.24 &                1 &                 64.86 &                1 &                 91.67 &               1 &                89.39 &            1 &             65.49 &              1 &               72.24 \\
G2&\includegraphics[scale=0.3]{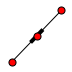}  &                0 &                   0   &                3 &                  4    &               2 &                 5.35 &                 4 &                   3.65 &                4 &                  3.57 &              2 &                5.57 &                5 &                  2.16 &                0 &                  0    &               3 &                 3.03 &            2 &              4.25 &              2 &                4.34 \\
G3&\includegraphics[scale=0.3]{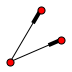}  &                0 &                   0   &                2 &                  5.33 &               4 &                 3.92 &                 3 &                   3.77 &                2 &                  4.58 &              3 &                4.45 &                2 &                  4.86 &                2 &                  2.78 &               2 &                 3.46 &            3 &              4.15 &              3 &                3.97 \\
G4&\includegraphics[scale=0.3]{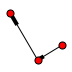}  &                2 &                  12.5 &                4 &                  3.33 &               3 &                 4.28 &                 2 &                   4.01 &                3 &                  3.57 &              4 &                2.9  &                4 &                  2.16 &                3 &                  1.85 &               5 &                 0.87 &            4 &              2.7  &              4 &                2.6  \\
G5&\includegraphics[scale=0.3]{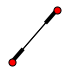}  &                0 &                   0   &                8 &                  1.33 &               9 &                 0.53 &                 6 &                   1.22 &                7 &                  1.23 &             11 &                0.67 &                3 &                  3.78 &                4 &                  1.85 &               4 &                 1.3  &            5 &              1.69 &              5 &                1.99 \\
G6&\includegraphics[scale=0.3]{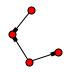}  &                0 &                   0   &                5 &                  3.33 &               7 &                 1.07 &                 5 &                   1.7  &                8 &                  1.23 &             90 &                0.22 &               32 &                  0.54 &                5 &                  0.93 &               0 &                 0    &            7 &              1.06 &              7 &                0.82 \\
G7&\includegraphics[scale=0.3]{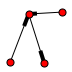}  &                0 &                   0   &                9 &                  1    &               0 &                 0    &                 9 &                   0.73 &               11 &                  0.56 &              7 &                1.34 &                6 &                  1.62 &                6 &                  0.93 &              11 &                 0.22 &            6 &              1.11 &              8 &                0.73 \\
G8&\includegraphics[scale=0.3]{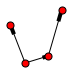}  &                0 &                   0   &                7 &                  1.67 &               8 &                 0.71 &                 8 &                   0.85 &                6 &                  1.34 &              6 &                1.56 &                0 &                  0    &                0 &                  0    &               0 &                 0    &            8 &              1.02 &             10 &                0.52 \\
G9&\includegraphics[scale=0.3]{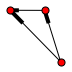}  &                0 &                   0   &                6 &                  3    &               5 &                 1.6  &                18 &                   0.36 &                5 &                  1.79 &              9 &                1.11 &                7 &                  1.62 &                0 &                  0    &               6 &                 0.65 &           10 &              0.67 &              9 &                0.61 \\
G10&\includegraphics[scale=0.3]{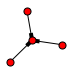} &                0 &                   0   &                0 &                  0    &              19 &                 0.36 &                13 &                   0.61 &               10 &                  0.67 &              8 &                1.34 &                0 &                  0    &                0 &                  0    &               0 &                 0    &            9 &              0.69 &              6 &                0.86 \\
G11&\includegraphics[scale=0.3]{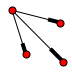} &                0 &                   0   &               10 &                  1    &               6 &                 1.25 &                10 &                   0.73 &                9 &                  0.78 &              5 &                1.78 &                8 &                  1.62 &                0 &                  0    &               0 &                 0    &           12 &              0.55 &             12 &                0.43 \\
G12&\includegraphics[scale=0.3]{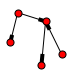} &                0 &                   0   &               12 &                  0.67 &              88 &                 0.18 &                 7 &                   0.97 &               13 &                  0.56 &             10 &                1.11 &               26 &                  0.54 &                0 &                  0    &               0 &                 0    &           11 &              0.58 &             11 &                0.49 \\
G13&\includegraphics[scale=0.3]{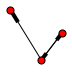} &                0 &                   0   &               18 &                  0.67 &              20 &                 0.36 &               187 &                   0.12 &               26 &                  0.22 &              0 &                0    &               35 &                  0.54 &                0 &                  0    &               9 &                 0.22 &           13 &              0.25 &             13 &                0.34 \\
G14&\includegraphics[scale=0.3]{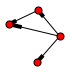} &                0 &                   0   &               15 &                  0.67 &              13 &                 0.36 &                17 &                   0.36 &               12 &                  0.56 &             13 &                0.45 &               29 &                  0.54 &                0 &                  0    &               0 &                 0    &           14 &              0.23 &             84 &                0.03 \\
G15&\includegraphics[scale=0.3]{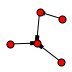} &                0 &                   0   &               47 &                  0.33 &               0 &                 0    &                16 &                   0.49 &               21 &                  0.22 &              0 &                0    &               10 &                  1.08 &                0 &                  0    &               0 &                 0    &           21 &              0.19 &             15 &                0.27 \\
G16&\includegraphics[scale=0.3]{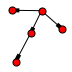} &                0 &                   0   &               54 &                  0.33 &              37 &                 0.18 &                11 &                   0.61 &               14 &                  0.45 &             29 &                0.22 &                0 &                  0    &                0 &                  0    &               0 &                 0    &           20 &              0.19 &             20 &                0.12 \\
G17&\includegraphics[scale=0.3]{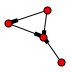} &                0 &                   0   &               17 &                  0.67 &              11 &                 0.53 &                21 &                   0.24 &               22 &                  0.22 &             67 &                0.22 &                0 &                  0    &                0 &                  0    &               0 &                 0    &           23 &              0.14 &             16 &                0.21 \\
G18&\includegraphics[scale=0.3]{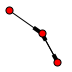} &                0 &                   0   &                0 &                  0    &               0 &                 0    &                80 &                   0.12 &                0 &                  0    &             12 &                0.67 &                0 &                  0    &                0 &                  0    &               0 &                 0    &           18 &              0.19 &             14 &                0.27 \\
G19&\includegraphics[scale=0.3]{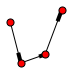} &                0 &                   0   &                0 &                  0    &               0 &                 0    &                12 &                   0.61 &               58 &                  0.11 &              0 &                0    &                9 &                  1.08 &                0 &                  0    &               0 &                 0    &           16 &              0.21 &             36 &                0.06 \\
G20&\includegraphics[scale=0.3]{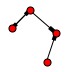} &                0 &                   0   &                0 &                  0    &              21 &                 0.36 &                30 &                   0.24 &              160 &                  0.11 &             99 &                0.22 &                0 &                  0    &                0 &                  0    &               0 &                 0    &           19 &              0.19 &             23 &                0.12 \\
\hline
\end{tabular}}
\begin{tablenotes}[normal,flushleft]
	\scriptsize
	\item \emph{Notes}: R indicates the rank of the cascade topology for the city, while FP indicates the frequency percentile

\end{tablenotes}
\caption{Frequent Cascades across different cities\label{tab:frequent}}

\end{table*}

\subsection*{Frequent Cascade Topologies}
\begin{figure*}[!ht]
	\centering
	\begin{minipage}{1.3in}
		\includegraphics[width=\linewidth]{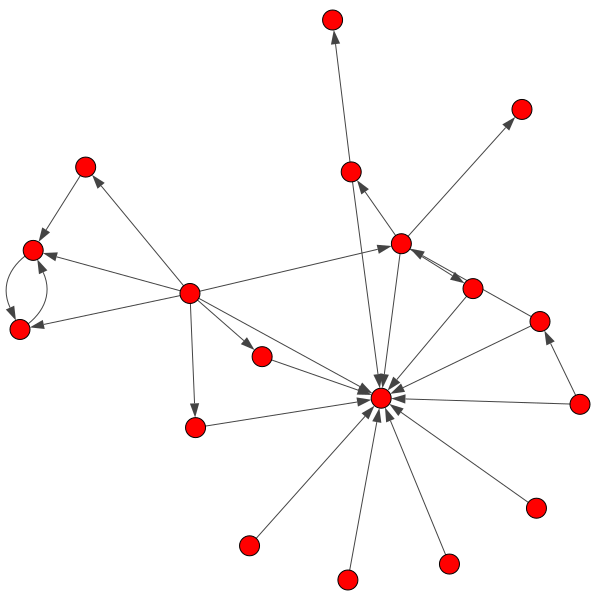}
		\subcaption{Urbana Champaign}
	\end{minipage}
	\begin{minipage}{1.3in}
		\includegraphics[width=\linewidth]{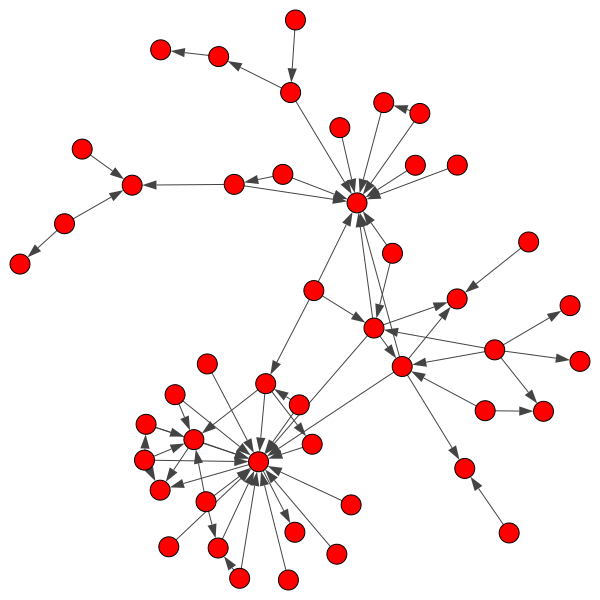}
		\subcaption{Charlotte}
	\end{minipage}
	\begin{minipage}{1.3in}
		\includegraphics[width=\linewidth]{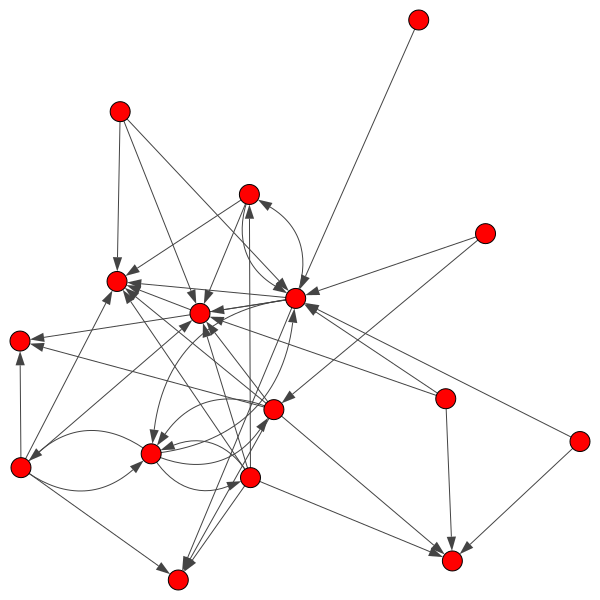}
		\subcaption{Edinburgh}
	\end{minipage}
	\begin{minipage}{1.3in}
		\includegraphics[width=\linewidth]{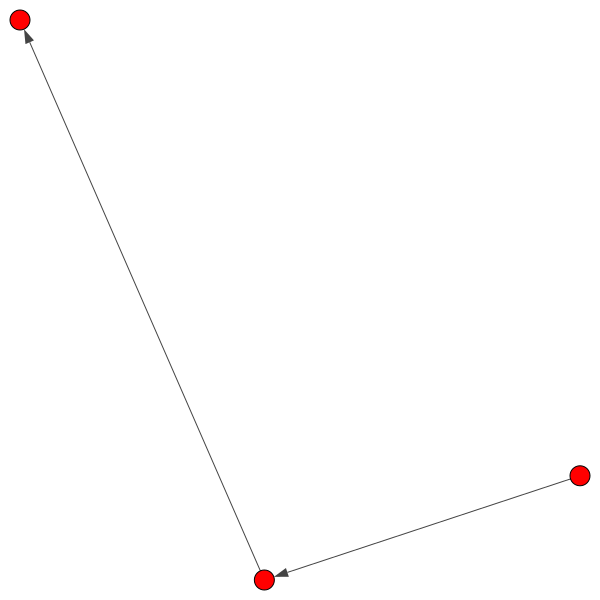}
		\subcaption{Karlsruhe}
	\end{minipage}
	\begin{minipage}{1.3in}
		\includegraphics[width=\linewidth]{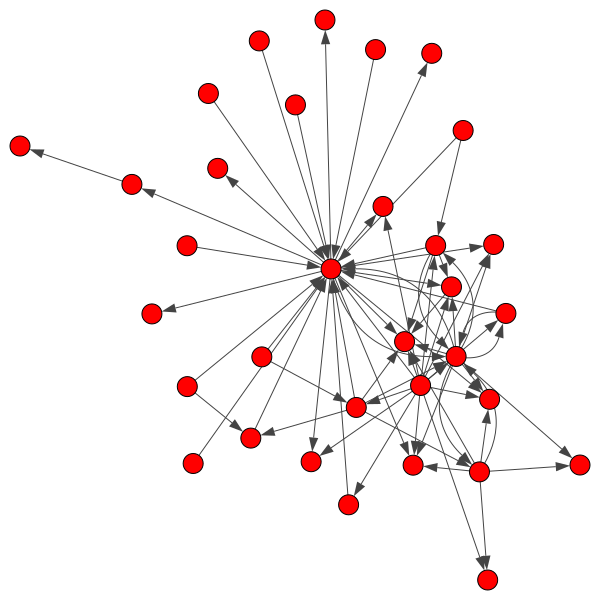}
		\subcaption{Madison}
	\end{minipage}
	\begin{minipage}{1.3in}
		\includegraphics[width=\linewidth]{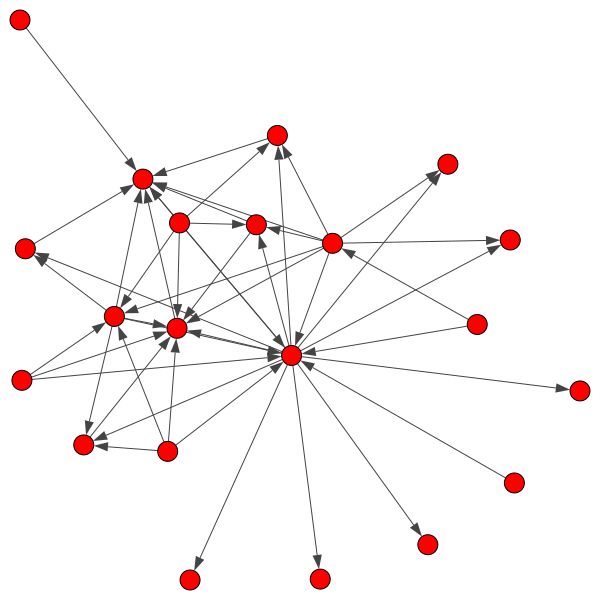}
		\subcaption{Montreal}
	\end{minipage}
	\begin{minipage}{1.3in}
		\includegraphics[width=\linewidth]{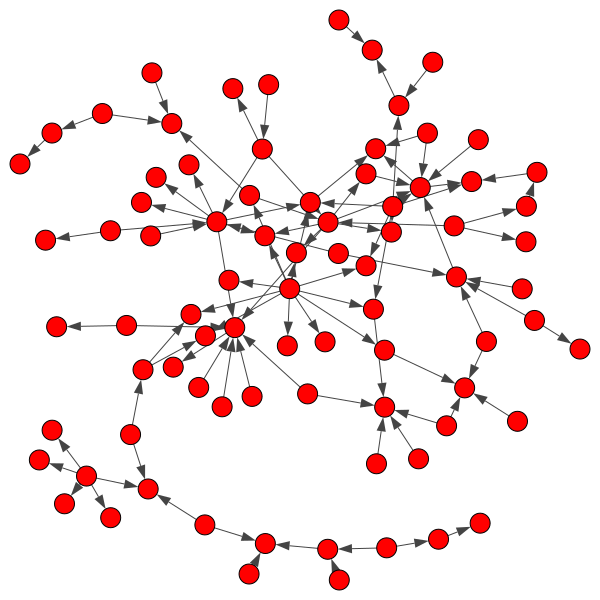}
		\subcaption{Pheonix}
	\end{minipage}
	\begin{minipage}{1.3in}
		\includegraphics[width=\linewidth]{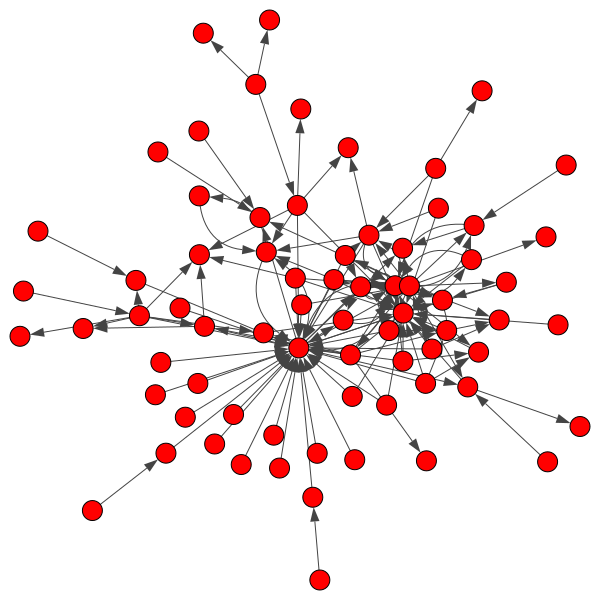}
		\subcaption{Pittsburgh}
	\end{minipage}
	\begin{minipage}{1.3in}
		\includegraphics[width=\linewidth]{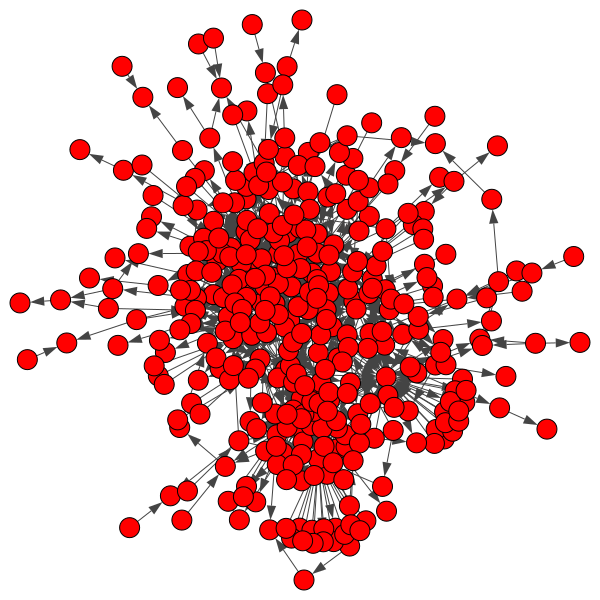}
		\subcaption{Las Vegas}
	\end{minipage}
	\begin{minipage}{1.3in}
		\includegraphics[width=\linewidth]{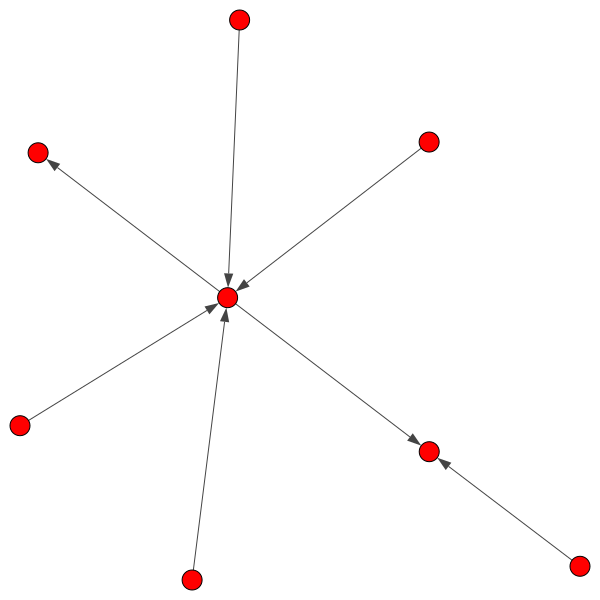}
		\subcaption{Waterloo}
	\end{minipage}
	\caption{Longest Cascades across cities\label{fig:max_cascade}}
\end{figure*}
As the next step in the structural analysis of cascades we wanted to see that what kind of cascades frequently arise in Yelp reviews across different cities. As discussed in \cite{leskovec:patterns_recnet:2006}, no polynomial time graph isomorphism algorithms exist and for the purpose of enumerating graphs one has to generate an efficiently computable signature of a graph and count the number of graphs with similar signatures. \cite{leskovec:patterns_recnet:2006} have proposed a multiple level signature for frequent subgraph mining. The signature used for hashing the graphs consists of a) Number of Nodes in the graph;  b) Number of edges in the graph;  c) Sorted in degree sequence of each of the node; and d) Sorted out degree sequence of each node. Using this technique we count the frequency of frequent cascades for each of the cities. Rank and frequency of the frequent cascade topologies for each of the cities is shown in the Table \ref{tab:frequent}.
As expected, the most common cascade i.e. G1 as shown in the Table \ref{tab:frequent} corresponds to the case in which a person participates in a cascade under the influence of only one other node. This corresponds to more than fifty percent of the cascades across all the cities and in some cases this particular type of cascade represents more than eighty percent of all the cascades of a city (e.g. Karlsruhe).
Comparing G2 and G10, we can see that receiving the recommendation or influence by more than once channel increases the likelihood of participation but at the same time, G3 and G11 shows that the presence of ``influential`` nodes also makes cascades longer.   
The cascades like G5 occur when both the user and the friend f either review or write tip for the same business id on the same day.
\subsection*{Longest Cascades}
On a similar line of exploration, we also visualize the longest cascade in each of the cities. The result of this exploration are shown in the Figure \ref{fig:max_cascade}. The presence of some influential central nodes is obvious in the cities like Las Vegas, Phoenix and Charlotte, the cities having higher number of business and reviewers. One clear exception to this trend is Waterloo where the size of the longest cascade is lower (90th percentile of the cascade length for each of the cities is shown in the table \ref{tab:sumstats}) as compared to the other cities but the number of business and reviewers on Yelp from waterloo is quite high

\section{Predicting Cascade Growth}
\label{sec:methods}
One of our main aim in this research was to see whether we can predict the growth of the cascade using a few initial reviews and tips. And as a second step, we wanted to analyze the features having high predictive power when it comes to predicting growth of cascades.
\subsection{Experimental Design}
Before going into the details of the feature engineering and cascades modeling, we want to specify the setup of our experiments in detail. 
First of all, we are trying to analyze the influence of a review or a tip made by the user on the activities of the members of the social network of that person. Though all of the yelp reviews are public and one can be influenced by the reviews and tips of the persons not in the friends list of that person, for the purpose of this research we consider only the friends of a person as a part of his social network. Yelp app and website has a specific interface of adding friends through Yelp, email or Facebook friends list \footnote{\url{https://www.yelp-support.com/article/How-do-I-add-friends-on-Yelp?l=en_US}}.\\
Secondly, the temporal ordering of reviews and tips by the users for a particular business gives us the edges of the cascades
The formulation of cascades as a prediction problem needs special attention because most of the cascades are small. \cite{cheng:be_predicted:2014} have summarized some of the issues and solution while modeling the growth of cascades. For the purpose of this study, we divide the cascades in each of the cities in two categories namely short cascades and big cascades where the cascades are categorized as big cascades if their length if greater than ninety percentile of the lengths of all the cascades of that city. The decision of ninetieth percentile may seem adhoc but it made sure that we are able to emphasize on the truly long cascades in a city. However, one impact of this decision was that we were having very few number of long cascades for Karlsruhe, Cleveland and Waterloo, hence we did not use the cascades from these cities in our predictive experiments. \\
The classification of cascades on the basis of percentiles may seem arbitrary but it makes sure that we are focusing on truly long cascades in each of the cities as the cascades in all of the cities are heavy tailed \ref{fig:dist}. A better experimental design would have been to model the growth of the cascades over time i.e. predicting whether the cascade would reach $2*k$ users, given that it has reached $k$ users(just like \cite{cheng:be_predicted:2014}) but our approach still sufficiently handles our objectives i.e. the study of predictability of cascades status ( long vs short) and the features determining the cascades length. \\
Furthermore, equal number of long and short cascades were selected for each of the city. 
\begin{table*}[!b]\centering
	\begin{threeparttable}

		\footnotesize
		\begin{tabular}{ll}
			\addlinespace
			\toprule
			\multirow{2}{*}{\textbf{Feature}} & \multirow{2}{*}{\textbf{Description}} \\
			\\
			\midrule
			&\textbf{Root (Original Poster) Features}\\
			Reviews\_Root & Number of reviews of the root node \\
			Friends\_Root  & Size of the social network of the root node\\
			YelpAge\_Root & The number of days on yelp\\
			TimeDelta\_Root & Average time difference between reviews\\
			Gender & Gender of the root node\\
			PositivityRatio\_Root & The number of positive reviews over the number of negative reviews for the root node\\
			Avg\{Pos,Neg\}Reviews\_Root & Average number of positive reviews by the root node\\
			StdReviews\_Root & Standard Deviation of reviews by the root node\\
			Fans\_Root & The number of fans of the root node\\
			Votes\{useful, funny, cool\} & The number of cool, funny or useful vote by the user\\
			Compliments\{hot,cute,cool,funny,writer,photos etc.\} & The number of compliments of a particular type \\
			EliteCount\_Root & The number of years the user was an elite\\
			
			\midrule
			&\textbf{Non Root Node Features}\\
			Friends\_NonRoot  & Average number of friends of non root nodes\\
			Reviews\_NonRoot  & Average number of reviews made by non root nodes\\
			YelpingAge\_NonRoot & Average yelping age of non root nodes\\
			TimeDelta\_NonRoot & Average time difference between reviews for the non root nodes\\
			MaleReviewers\_NonRoot & The ratio of number of male non root reviewers and total number of non root reviewers\\
			PositivityRatio\_NonRoot & The average of the average positivity ratio of all the reviews by the non root nodes\\
			PositiveReviews\_NonRoot & Average of average of all the positive reviews by non root nodes\\
			NegativeReviews\_NonRoot & Average of average of all the negative reviews by non root nodes\\
			Reviews\_NonRoot & Average and Standard Deviation of Standard Deviation of reviews of non root nodes\\
			Votes\{useful, funny, cool\}\_NonRoot & Average and Standard Deviation of the number of cool, funny or useful vote by the non root nodes\\
			EliteCount\_NonRoot & Average and Standard Deviation of the number of years the non root node users have been elite\\
			Compliments \{hot,cute,cool,funny,writer,photos etc.\}&Average and standard deviation for the number of compliments of specific type\\
			\midrule
			&\textbf{Root Review Features}\\
			Funny, Useful and Cool Votes  & The number of cool useful and funny votes for the first review\\
			Positivity & Positive of the root review\\
			\midrule
			&\textbf{Non Root Review Features}\\
			Funny, Useful and Cool Votes  & The total number and average of funny useful and cool votes for the non root reviews\\
			Positivity & Average Positive of the non root reviews\\
			\midrule
			&\textbf{Business Features}\\
			Rating  & Average rating stars for the business\\
			No of reviews and votes & Number of reviews and votes for the business\\
			\bottomrule
		\end{tabular}
		\caption{Features used for modeling short vs long cascades\label{tab:features}}
	\end{threeparttable}
\end{table*}

\subsection{Features Engineering}
\label{feat_eng}

The table \ref{tab:features} shows the list of features used for predicting longevity of the cascades. Primarily the features are divided into five main categories namely:Business related features, Root node features, Non-root node features, Root review features, Non-root review features. 
\subsection{Classification and Model Selection}
\label{sec:models}

Using the features specified in the last section, we used a variety of supervised learning algorithms to classify the cascades as long or short cascades. 
Since our data has a large number of features relative to observations, we focus on learners that are robust to over-fitting, such as regularized logistic regression \cite{zou:regularization:2005}, gradient boosting \cite{friedman:greedy:2001}. Performance was comparable across these classifiers, although as expected these methods generally performed better than unregularized alternatives. To streamline the analysis summarized in the section \ref{sec:results}, we report only the results from gradient boosting, which outperformed the other classifiers by a small margin. In all cases, we report the average accuracy across testing sets from 5-fold cross validation.

The level of a feature used as a decision node in the decision tree can be used to assess the relative importance of the feature to predict the longevity of the cascades. We use the feature importance calculated in this way to identify important feature categories as described in the section \ref{sec:results}.

\begin{figure*}[!h]
	\centering
	\begin{subfigure}{0.45\linewidth}        
		\centering
		\includegraphics[width=\linewidth]{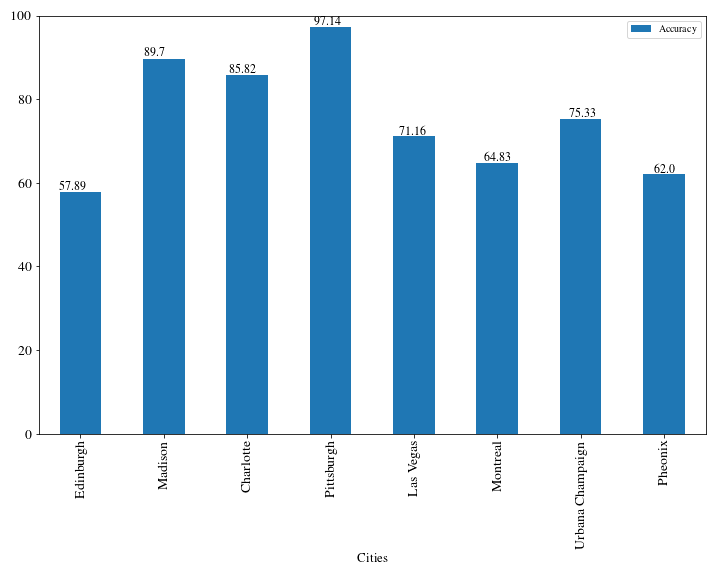}
		\caption{Predictive Accuracy}
		\label{fig:accuracy}
	\end{subfigure}
	\begin{subfigure}{0.45\linewidth}        
		\centering
		\includegraphics[width=\linewidth]{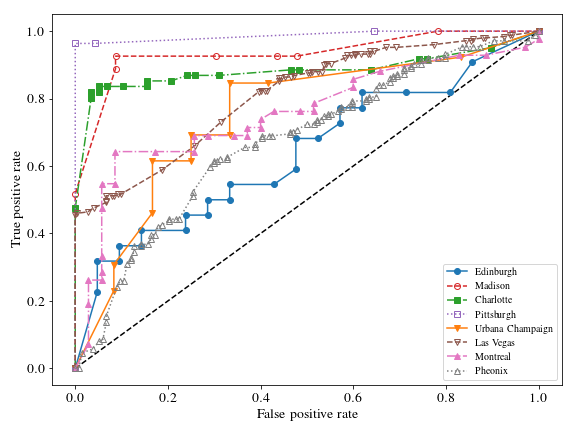}
		\caption{ROC Curve\label{fig:roc}}
	\end{subfigure}
	\caption{Results for cascade prediction;}
	\label{fig:roc_curve}
\end{figure*}
\begin{table*}[!h]\centering
	\begin{threeparttable}

		\footnotesize
		\begin{tabular}{lllc}
			\addlinespace
			\toprule
			\multirow{2}{*}{\textbf{Feature}} & \multirow{2}{*}{\textbf{Most Important Feature}} & \multirow{2}{*}{\textbf{Second Most Important Feature}} & \multirow{2}{*}{\textbf{Predictive Accuracy (\%)}}\\
			\\
			\midrule

			\textbf{Edinburgh} & Variance in the yelping age of the non root reviewers&No of useful votes for the root reviews&57.89\\
			\textbf{Madison}  & Variance in the yelping age of the non root reviewers&No of hot compliments for the business&89.7\\
			\textbf{Charlotte} & Variance in the yelping age of the non root reviewers&Average star rating for the non root node&85.82\\
			\textbf{Pittsburgh} & Variance in the yelping age of the non root reviewers&No of photo compliments for the business&97.14\\
			\textbf{Urbana Champaign} & Average number of elite ratings for the non root reviewers&No of useful votes on the reviews of the business&75.33\\
			\textbf{Las Vegas} & Variance in the yelping age of the non root reviewers&No of useful votes on the reviews of the business&71.16\\
			\textbf{Phoenix} & Number of cool votes for the business&Average star rating for the non-root reviews&62.0\\
			\textbf{Montreal} & Number of stars for the root node&No of cool compliments for the root node&64.83\\
			
			\bottomrule
		\end{tabular}

		\caption{Top predictive features\label{tab:top_features}}
	\end{threeparttable}
\end{table*}

\section{Results and Discussion}
\label{sec:results}

As stated earlier, our first goal in this research was to evaluate whether cascades growth can be predicted on Yelp or not. To this end, he predictive accuracy and ROC curve ( based on our best performing gradient boosting algorithm as described in the section \ref{sec:methods}) for predicting  status (long vs short) of cascades across different cities is shown in the Figures \ref{fig:accuracy} and \ref{fig:roc} respectively.


In addition to calculating the predictive accuracy and ROC for the cascades in each of the cities, we also determine the importance of different type of features for each of the city as well. We have generated five different type of feature categories explained as follows.
\begin{itemize}
	\item \textbf{Root Features:} All the features of the original node(i.e. person) who started the cascade are categorized as root node node features. Some example of the root node review features include the gender of the root node and the number of reviews the root node has contributed
	\item \textbf{Non Root Features:} This category includes the features of all the non root nodes that contribute to the cascade. Some example of the features in this category include average number of friends for the non root nodes for a cascade, average number of reviews by all the non root nodes.
	\item \textbf{Root Review Features:} All the characteristics of the text of the root review ( the review that started the cascade) are categorized as the root review features. For example, the number of ``cool'' votes received by the root review.
	\item \textbf{Non Root Review Features:} All the characteristics of the non root reviews contributing to the cascade are categorized under the non root review features. Total number of cool votes received by all the non root reviews is one such example
	\item \textbf{Business Features:} The last category of features includes all the features for the business receiving the review or tips. Total number of reviews received by the node is one example of business related features.
\end{itemize}
\begin{figure*}[!t]
	\begin{minipage}[h]{1.7in}
		\includegraphics[width=\linewidth]{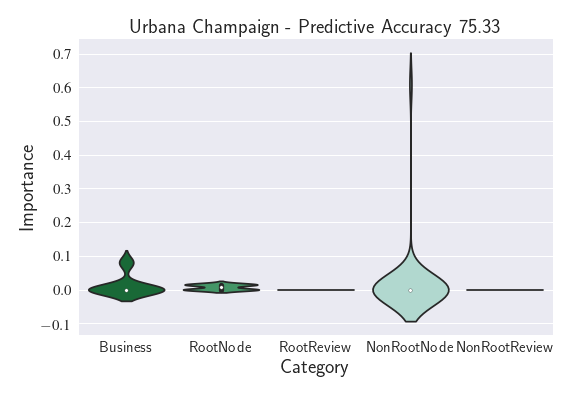}
	\end{minipage}
	\begin{minipage}[h]{1.7in}
		\includegraphics[width=\linewidth]{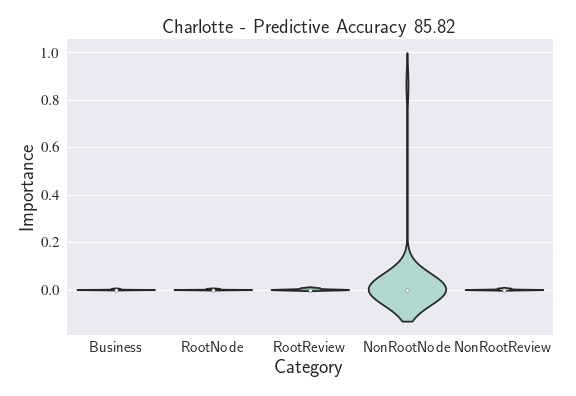}
	\end{minipage}
	\begin{minipage}[h]{1.7in}
		\includegraphics[width=\linewidth]{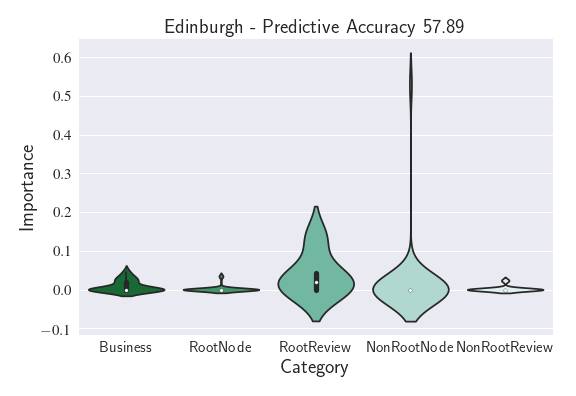}
	\end{minipage}
	\begin{minipage}[h]{1.7in}
		\includegraphics[width=\linewidth]{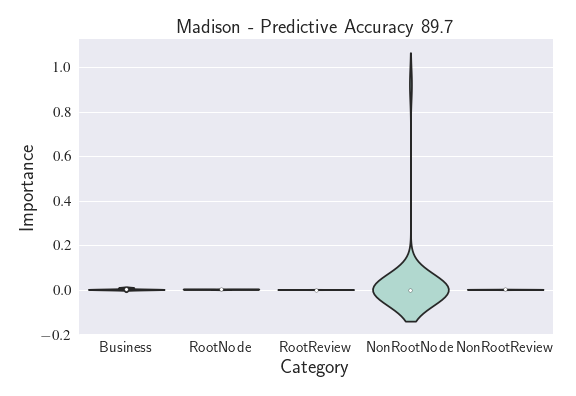}
	\end{minipage}
	\begin{minipage}[h]{1.7in}
		\includegraphics[width=\linewidth]{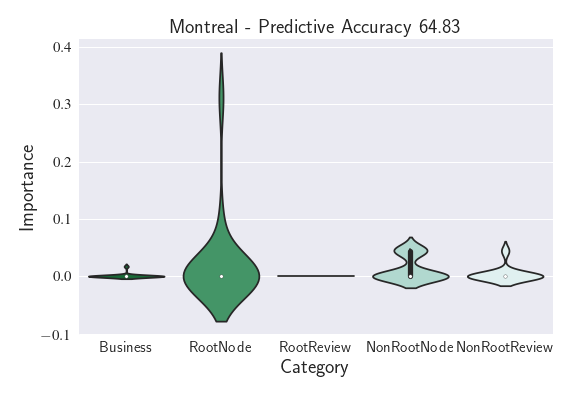}
	\end{minipage}
	\begin{minipage}[h]{1.7in}
		\includegraphics[width=\linewidth]{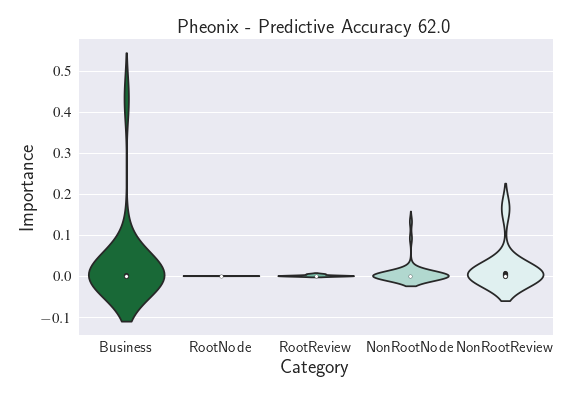}
	\end{minipage}
	\begin{minipage}[h]{1.7in}
		\includegraphics[width=\linewidth]{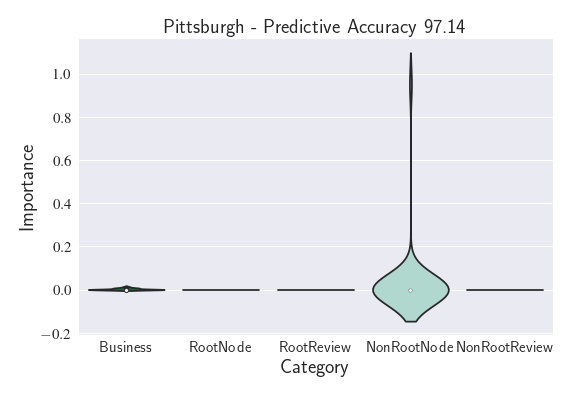}
	\end{minipage}
	\begin{minipage}[h]{1.7in}
		\includegraphics[width=\linewidth]{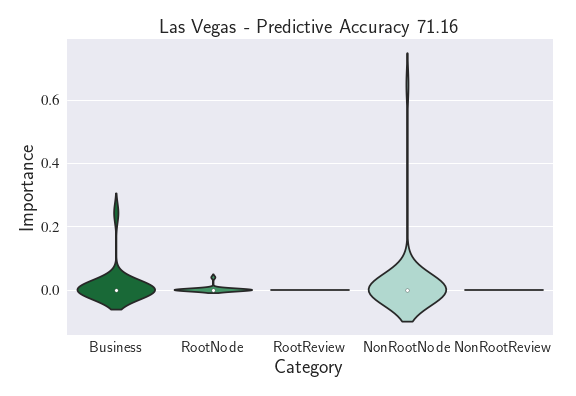}
	\end{minipage}	
	\caption{Importance of features across cities\label{fig:violin}}
\end{figure*}

More details of different features used in the analysis can be seen in the Table \ref{tab:features}. The sentiment prediction(positivity) for the actual text of the reviews is performed using a model pre-trained on the Yelp and Amazon reviews with ratings greater than being considered as positive reviews.
Importance for different type of features is shown as violin plots \cite{hintze1998violin} in the Figure \ref{fig:violin}. To understand which features are more important for prediction of cascades, we calculate the importance of each feature using gradient boosting classiﬁer. The importance of features is the average of importance in each of the individual trees in the gradient boosting. Each of the trees calculates the importance of features using the information gain with intent to favor pure subtrees at each level of the tree.

Some interesting patterns can be observed in the results shown in the Figures \ref{fig:roc_curve},\ref{fig:violin}. For instance, cascades of the cities like Phoenix, Montreal and Edinburgh cannot be predicted very accurately while those of Pittsburgh and Madison can be predicted with quite high accuracy.

In all the cities, where the cascades can be predicted with higher accuracy, the features belonging to the non root nodes have higher predictive importance. Amongst the cities with low predictive results, only Edinburgh has got higher feature importance for the non root node category. In none of the cities where the cascades can be predicted with higher accuracy, the root nodes and root reviews have higher predictive importance which shows that the first reviewer and the first review may not be that importance in the case of cascades in Yelp reviews or in other words, non influential nodes may start the cascades.

Top two features in each of the cities are shown in the Table \ref{tab:top_features}.  The variance in the yelping age of the non root reviewers is the top feature in most of the cities which indicates that the users who are active and relatively long term users of Yelp are more likely to be part of the cascades. Different type of votes that a review, reviewer or a business may receive is the dominant type of second most important feature.
%

\section{Conclusions}
\label{sec:conclusion}
In this paper, we have tried to analyze the cascades on Yelp in terms of the predictability of them being long or short. At the same time, we have tried to identify the important properties of these cascades and the properties of the nodes and the business involved in these cascades. Though, the predictive performance of our models vary across different cities but some notable conclusions can be drawn from our analysis. First of all, in most of the cities with the higher number of business and reviews the longevity of the cascades can be predicted quite accurately. Secondly, the characteristics of the non-root reviewers are a more powerful indicator of the popularity of the cascade as compared to other categories like root reviewer and root review characteristics. Lastly, the variance in the age of the non root node reviewers is the most important feature in most of the cities.

There have been very few attempts to systematically analyze the information flow or propagation on Yelp but these information cascades have the potential to help business owners and Yelp both to improve their services. Most of the focus of this research has been to identify the structural and nonstructural features of the cascades on Yelp and see the relationship between these  features and length of the cascades. In future, we want to incorporate additional data sources like pictures and images of the businesses in our analysis. We also want to experiment with the recent deep learning based methods (e.g. \cite{li2017deepcas}) to model the growth of cascades.  

Right now, we are not imposing any time bound on the edges of the cascades which may imply that the successive edges in a cascade may be months apart. In our future work, we want to apply additional constraint of time bound on the successive edges in a cascade. Though our focus in this work has been on the cascades in Yelp social networks but a similar analysis can be extended to other types of social network provided significant data is available.

\pagebreak

\bibliographystyle{ACM-Reference-Format}
\bibliography{yelp-cascades-references} 

\end{document}